# A Matrix Laurent Series-based Fast Fourier Transform for Blocklengths $N \equiv 4 \pmod{8}$

H.M. de Oliveira, R.M. Campello de Souza and R.C. de Oliveira

*Abstract*— General guidelines for a new fast computation of blocklength $8m+4$ DFTs are presented, which is based on a Laurent series involving matrices. Results of non-trivial real multiplicative complexity are presented for blocklengths $N \leq 64$, achieving lower multiplication counts than previously published FFTs. A detailed description for the cases $m=1$ and $m=2$ is presented.

*Index Terms*— fast algorithms, FFT, Laurent series, Heideman bound.

## I. INTRODUCTION

FOURIER transforms have been playing a major role in quite a lot of areas, especially in the fields linked to Signal Processing [1, 2]. In practical cases, its evaluation is not carried out analytically, but rather numerically and, in most times, there is no analytical expression of the signal to be analyzed. The successful application of transform techniques is mainly due to the existence of the so-called fast algorithms [3, 4]. Therefore, techniques for computing discrete transforms with a low multiplicative complexity, have been an object of interest for a long time.

This paper proposes a new fast algorithm for computing the Discrete Fourier Transform (DFT) of sequences of particular lengths $N$, namely those for which $N \equiv 4 \pmod{8}$, but can also be extended for $N \equiv 0 \pmod 4$.

Let $N$ be the number of time-domain samples of a sequence $v = (v_n)$, $n=0,1,2,...,N$-1. The DFT of $v$ is given by the sequence $V = (V_k)$, of length $N$, in the frequency domain, defined by

$$V_k := \sum_{n=0}^{N-1} v_n \exp\left(-\frac{j2\pi kn}{N}\right). \quad (1)$$

In 1965, J.W. Cooley and J.W. Tukey introduced a revolutionary idea which later became known as the fast Fourier transform (FFT) [5]. The FFT is a milestone in the theory of algorithms [6], and more specifically in the Signal Processing field [7, 8].

With the advent of VLSI and the development of the DSP (Digital Signal Processor, processors in *chip*) to implement signal processing techniques, the DFT became the most attractive tool for spectrum evaluation [9-11]. The cost reduction of DSPs and the astonishing capacity achieved by up to date processors (e.g., dozens of GFlops–Giga floating-point operations per second–and TFlops) [13], together with novel and efficient signal processing techniques, is turning real-time application feasible for several kind of signals. Therefore, discrete transforms become the widespread tool in spectral analysis [14].

There are several standard FFT algorithms in the literature, including Cooley-Tukey, Good-Thomas and the Winograd-Fourier algorithm (WFTA) [15]. A lucid tutorial review of fast Fourier techniques by Duhamel and Vetterli is available in [16].

In 1987, Heideman investigated the arithmetical complexity of the DFT and derived lower bounds on the multiplicative complexity for computing it [17]. Let $\mu_{DFT}(N)$ be the minimal multiplicative complexity of the exact computation of a blocklength $N$ DFT.

**Theorem** (Heideman). For $N = \prod_{i=1}^{m} p_i^{e_i}$, where $p_i$, $i=1,...,m$ are distinct primes and $e_i$, $i=1,...,m$ are positive integers, it follows that

$$\mu_{DFT}(N) = 2N - \sum_{i_1=0}^{e_1}\sum_{i_2=0}^{e_2}\cdots\sum_{i_3=0}^{e_m} \phi\left(\gcd\left(\prod_{j=1}^{m} p_j^{i_j}, 4\right)\right).$$

$$\left(1 + \sum_{d_1 | \frac{\phi(p_1^{i_1})}{\phi(\gcd(p_1^{i_1},4))}} \sum_{d_2 | \frac{\phi(p_2^{i_2})}{\phi(\gcd(p_2^{i_2},4))}} \cdots \sum_{d_m | \frac{\phi(p_m^{i_m})}{\phi(\gcd(p_m^{i_m},4))}} \frac{\prod_{k=1}^{m}\phi(d_k)}{\phi(lcm(d_1,d_2,...,d_m))}\right)$$

where $\phi(.)$ is the Euler totient function, $gcd(.,.)$ denotes the greatest common divisor and $lcm(.,.)$ the least common multiple.

**Proof**. See [17, M.T. Heideman, p.98]. ∎

This proof is based on evaluating the multiplicative complexity of a set of polynomial products.

In 2000, de Oliveira, Cintra, Campello de Souza [18, 19], introduced an algorithm based on multilayer decomposition to calculate the DFT via the discrete Hartley transform (DHT), which meets the minimal complexity for blocklengths up to $N$=24 [20].

Another approach to quantize a Fourier series consists of digitalizing the basis of signals used in the decomposition [21]. Matrices so derived are always "*quasi*-diagonal matrices" and their inverses are "*quasi*-identity matrices". The Möbius inversion formula was also used to derive guesstimates of the DFT from the coefficients of the quantized series. The analysis is rather similar to the one proposed by Cintra and col. in the framework of rounded Hartley transform [22, 23].

## II. DFT AS A MATRIX LAURENT SERIES

The first step towards the FFT proposed in this paper is to rewrite eqn(1) in matrix form:

H.M de Oliveira (qPGOM) and R.M. Campello de Souza are with the Signal Processing Group, Federal University of Pernambuco (UFPE), C.P. 7800, CEP: 50711-970, Recife-PE; R.C. de Oliveira is with Amazon State University, (UEA), Av. Darcy Vargas, 1200 Parque 10 - CEP 69065-020 Manaus-AM. (e-mail: rcorrea.oliveira@gmail.com, {hmo, ricardo} @ufpe.br).

$$\begin{bmatrix} V_0 \\ V_1 \\ V_2 \\ \vdots \\ V_{N-1} \end{bmatrix} = \begin{bmatrix} 1 & 1 & 1 & \ldots & 1 \\ 1 & W & W^2 & \ldots & W^{N-1} \\ 1 & W^2 & W^4 & \ldots & W^{2.(N-1)} \\ \vdots & \vdots & \vdots & \ddots & \vdots \\ 1 & W^{N-1} & W^{2(N-1)} & \ldots & W^{(N-1).(N-1)} \end{bmatrix} \begin{bmatrix} v_0 \\ v_1 \\ v_2 \\ \vdots \\ v_{N-1} \end{bmatrix},$$

or $(V_k) = [\text{DFT}] \cdot (v_n)$. Since $W := e^{-j\frac{2\pi}{N}}$ has order $N$, there are only $N$ distinct powers of $W$ in the set $\{W^0, W^1, W^2, W^3, \ldots, W^{(N-1).(N-1)}\}$.

This paper deals with blocklengths $N \equiv 4 \pmod 8$ so as to guarantee that there exists always a power of $W$ yielding the eigenvalues of the DFT, i.e., $\pm 1, \pm j$ [24]. These terms do not contribute to the multiplicative complexity, because

$W^0 = 1$, $W^{N/4} = -j$, $W^{N/2} = -1$ and $W^{3N/4} = j$. (2)

The exponents of $W$ in expression (2) generate a set of four points that lie on the real or imaginary axis. This fact is associated to the set

$C_0 := \{0, N/4, N/2, 3N/4\}$

and we are looking for particular symmetries in the four quadrants of the Argand-Gauss plane [25].

The set of exponents of the distinct powers of $W$, $\{W^0, W^1, W^2, \ldots, W^{(N-1)}\}$ (the $N$-th roots of unity), is then partitioned into $N/4$ (disjoint) classes (worth to remark that $4 \mid N$):

$C_m := \{x \in \mathbb{N} \cap [0,N) \mid 4x \equiv 4m \pmod N\}$, where $\mathbb{N}$ is the set of natural numbers and $m = 0, \pm 1, \pm 2, \ldots, \pm\left(\frac{N}{4}-1\right)/2$.

**Proposition 1**. The above mentioned classes $\{C_m\}$ engender a partition of the ensemble of integers $\{0,1,2,\ldots,N-1\}$, i.e., $\forall m \neq m'$, $C_m \cap C_{m'} = \varnothing$ and $\bigcup_{m=-(N/4-1)}^{N/4-1} C_m = \{0,1,2,\ldots,N-1\}$.

**Proof**. Suppose (by *reduction ad absurdum*) that there exists a pair $m \neq m'$ such that $C_m \cap C_{m'} \neq \varnothing$. Therefore, there is a common element $x \in C_m$ and $x \in C_{m'}$ such that $4x \equiv 4m$ $\pmod N$ and $4x \equiv 4m' \pmod N$. Therefore $4m \equiv 4m' \pmod N$, which is the same as $m \equiv m' \pmod{N/4}$, a contradiction. The cardinality of a set $C_m$ for each $m$ is $\|C_m\|=4$. There are $N/4$ disjoint classes, so that $\left\| \bigcup_{m=-(N/4-1)}^{N/4-1} C_m \right\| = 4 \cdot (N/4) = N$ and the classes $\{C_m\}$ form a partition of $\{0,1,2,\ldots,N-1\}$. ∎

For the sake of simplicity, we deal only with the matrix of exponents of $W$ in the DFT matrix. Let us define an $N \times N$ matrix $M := (kn \pmod N)$, whose elements belong to the set $\{0,1,2,\ldots,N-1\}$.

We also define an operator $\chi_l$ over an $N \times N$ matrix for each $l = 0,1,2,\ldots,N-1$, which yields a new $N \times N$ binary matrix whose elements are $(\delta_{l,m_{k,n}})$, where $\delta$ is the Kronecker symbol.

Finally, we define a matrix $M_m$ associated with each class $C_m$, for $m = 0, \pm 1, \pm 2, \ldots, \pm\left(\frac{N}{4}-1\right)/2$:

$$M_m := \sum_{l \in C_m} (j)^{l-m} \chi_l(M). \quad (3)$$

For instance, $m=0$ corresponds to the additive part of the DFT transform matrix:

$M_0 = 1 \cdot \chi_0(M) - j \cdot \chi_{N/4}(M) - 1 \cdot \chi_{N/2}(M) + j \cdot \chi_{3N/4}(M)$.

Consider a (possibly infinite) matrix $A$ expressed in terms of block matrices in the form:

$A = (\ldots, A_{-1}, A_0, A_1, \ldots)$,

where $A_l$ are $N \times N$ submatrices of $A$.

From the matrix $A$, the following formal power series is called Laurent series of the matrix $A$ [26, 27]:

$$A(z) := \sum_{l=-\infty}^{+\infty} A_l \cdot z^l.$$

Thus, $A(z)$ is a Laurent series with matrix coefficients. In particular cases where $A(z) := \sum_{l=N_1}^{N_2} A_l \cdot z^l$, $N_1, N_2 \in \mathbb{Z}$, then

$g := N_2 - N_1 + 1$

is the genus of $A(z)$ and $A$ [27].

Let us now name $\mathfrak{M}$ as the matrix associated with the submatrices $M_m$:

$$\mathfrak{M} := \left( M_{-\frac{N/4-1}{2}}, \ldots, M_{-1}, M_0, M_1, \ldots M_{\frac{N/4-1}{2}} \right).$$

The Laurent series of the matrix $\mathfrak{M}$ is

$$\mathfrak{M}(z) = M_{-(N/4-1)/2} \cdot \frac{1}{z^{(N/4-1)/2}} + \ldots$$
$$+ M_{-2} \cdot \frac{1}{z^2} + M_{-1} \cdot \frac{1}{z} + M_0 + M_1 \cdot z + M_2 \cdot z^2 + \ldots$$
$$+ M_{(N/4-1)/2} \cdot z^{(N/4-1)/2},$$

which has genus $g = N/4$ (in the filter bank framework, the notation of Laurent series is referred to as the polyphase representation [1]).

The evaluation of the discrete Fourier spectrum corresponds to a product of the discrete-time data sequence by the DFT transform matrix $[\text{DFT}] = \mathfrak{M}(z)\big|_{z=W}$, that is, the DFT transform matrix is

$$\mathfrak{M}(z)\big|_{z=W} = \sum_{-((N/4)-1)/2}^{+((N/4)-1)/2} M_m \cdot W^m. \quad (5)$$

Since the multiplications by $W^m$ and $W^{-m}=(W^m)^*$ for a fixed value of $m$ are essentially equivalent, these matrices can be combined by considering $\{M_{-m}, M_m\}$ and writing this coupled matrix in the standard echelon form (SEF, referred here as *rref*, *r*ow-*r*educed *e*chelon *f*orm as in Matlab and Mathcad software).

Steps of simplification consider only powers of:
$\{W^0, W^1, W^2, W^3, \ldots, W^{(N-1).(N-1)}\}$;
$\{W^0, W^1, W^2, W^3, \ldots, W^{(N-1)}\}$;
$\{W^0, W^1, W^{-1}, W^2, W^{-2}, \ldots, W^{(N/4-1)/2}, W^{-(N/4-1)/2}\}$;

$\Re e W = \cos\frac{2\pi}{N}$ and $\Im m W = \sin\frac{2\pi}{N}$.

For $N \equiv 0 \pmod 8$ there is a lack of symmetry, with more positive than negative terms in expression (5). For instance, for $N=8$, the decomposition takes the form: $\mathfrak{M}(z)\big|_{z=W} = M_0 + M_1 W$. For $N=16$ the algorithm yields $\mathfrak{M}(z)\big|_{z=W} = M_{-1}W + M_0 + M_1 W + M_2 W^2$. In general, the only asymmetric class in the series that is appended to the symmetric classes is $C_{N/8}$, $N \geq 8$.

## III. THE NEW FFT ALGORITHM.

The fast algorithm is written in terms of the matrices $\{M_m\}$ according with the following decompositions:

$$\Re e DFT = \left\{ \Re e(M_0) + \sum_{m=1}^{(N/4-1)/2} \Re e(M_m + M_{-m}) \cdot \cos\frac{2\pi m}{N} \right\}$$
$$+ \left\{ \sum_{k=1}^{(N/4-1)/2} \Im m(M_m - M_{-m}) \cdot \sin\frac{2\pi m}{N} \right\},$$
$$\Im m DFT = \left\{ \Im m(M_0) + \sum_{m=1}^{(N/4-1)/2} \Im m(M_m + M_{-m}) \cdot \cos\frac{2\pi m}{N} \right\}$$
$$- \left\{ \sum_{k=1}^{(N/4-1)/2} \Re e(M_m - M_{-m}) \cdot \sin\frac{2\pi m}{N} \right\}.$$

The matrices $\Re e(M_0)$ and $\Re e(M_m \pm M_{-m})$ are then written in SEF, as well as the corresponding matrices $\Im m(M_0)$ and $\Im m(M_m \pm M_{-m})$.

The multiplicative complexity of the fast transform can be computed by

$$\sum_{m=1}^{\left(\frac{N}{4}-1\right)/2} rank\begin{pmatrix} \Re e(M_m + M_{-m}) \\ \Im m(M_m + M_{-m}) \end{pmatrix} + rank\begin{pmatrix} \Re e(M_m - M_{-m}) \\ \Im m(M_m - M_{-m}) \end{pmatrix}. \quad (6)$$

In every case examined so far, no reduction of rank was achieved when stacking the matrices $\Re e(M_m \pm M_{-m})$ and $\Im m(M_m \pm M_{-m})$, and the multiplicative complexity of the FFT was always given by

$$2\sum_{m=1}^{\left(\frac{N}{4}-1\right)/2} rank(\Re e(M_m + M_{-m})) + rank(\Im m(M_m + M_{-m})).$$

In the naïve example $N=8$, there are only two matrices associated with the multiplicative terms, namely:

$$\Re e(M_1) = \Im m(M_1) = \begin{pmatrix} 0 & 1 & 0 & 0 & 0 & -1 & 0 & 0 \\ 0 & 0 & 0 & 1 & 0 & 0 & 0 & -1 \end{pmatrix},$$

so only two multiplications ( $\cos(\pi/4) = \sin(\pi/4)$ ) are required. It is worth to observe that the number of real multiplications is two unities less than that one computed by eqn.6 when $N \equiv 0 \pmod{4}$, because a multiplication by $\exp(j\pi/4)$ is included.

## IV. AN FFT FOR BLOCKLENGTH N=12

For $N=12$, we start gathering the elements of exponents in the class $\{0, 3, 6, 9\}$, which are not associated with multiplications (see eqn 1): this corresponds to the set $C_0$. The matrix $M$ with the exponents of the terms of the DFT matrix is

$$M := \begin{pmatrix} 0 & 0 & 0 & 0 & 0 & 0 & 0 & 0 & 0 & 0 & 0 & 0 \\ 0 & 1 & 2 & 3 & 4 & 5 & 6 & 7 & 8 & 9 & 10 & 11 \\ 0 & 2 & 4 & 6 & 8 & 10 & 0 & 2 & 4 & 6 & 8 & 10 \\ 0 & 3 & 6 & 9 & 0 & 3 & 6 & 9 & 0 & 3 & 6 & 9 \\ 0 & 4 & 8 & 0 & 4 & 8 & 0 & 4 & 8 & 0 & 4 & 8 \\ 0 & 5 & 10 & 3 & 8 & 1 & 6 & 11 & 4 & 9 & 2 & 7 \\ 0 & 6 & 0 & 6 & 0 & 6 & 0 & 6 & 0 & 6 & 0 & 6 \\ 0 & 7 & 2 & 9 & 4 & 11 & 6 & 1 & 8 & 3 & 10 & 5 \\ 0 & 8 & 4 & 0 & 8 & 4 & 0 & 8 & 4 & 0 & 8 & 4 \\ 0 & 9 & 6 & 3 & 0 & 9 & 6 & 3 & 0 & 9 & 6 & 3 \\ 0 & 10 & 8 & 6 & 4 & 2 & 0 & 10 & 8 & 6 & 4 & 2 \\ 0 & 11 & 10 & 9 & 8 & 7 & 6 & 5 & 4 & 3 & 2 & 1 \end{pmatrix}$$

There are only $N/4=3$ classes, namely:

$C_0=(0, 3, 6, 9)$,      $0 \equiv 24 \equiv 12 \equiv 36 \pmod{12}$
$C_1=(1, 4, 7, 10)$,      $4 \equiv 28 \equiv 16 \equiv 40 \pmod{12}$
$C_{-1}=(11, 2, 5, 8)$.    $44 \equiv 20 \equiv 8 \equiv 32 \pmod{12}$.

In this particular case, the greatest index is $(N/4-1)/2=1$. Indeed, $C_{-1}$, $C_0$, $C_1$ are a partition of $\{0,1,2,\ldots,11\}$, as expected.

It is straightforward to observe that given $C_0$, the elements of $C_1$ can be directly derived by adding 1 (mod $N$) to each element of $C_0$; $C_{-1}$ by subtracting 1 (mod $N$) to each element of $C_0$, and so on.

In order to clarify the approach, we take the set of powers of $W$:
$$\{1, W, W^2, W^3, W^4, W^5, W^6, W^7, W^8, W^9, W^{10}, W^{11}\}.$$

Since $W^0 = 1$, $W^3 = -j$, $W^6 = -1$, $W^9 = j$, the following classes are considered:

$C_0=\{0,3,6,9\} \Rightarrow 1, -j, -1, j$
$C_1=\{1,4,7,10\} \Rightarrow W^1=1.W, W^4=-j.W, W^7=-W, W^{10}=j.W.$
$C_{-1}=(11, 2, 5, 8) \Rightarrow W^{11}=W^*, W^2=-jW^*, W^5=-W^*, W^8=j.W^*.$

The operations involving product by the eigenvalues (elements of $C_0$) and/or the conjugacy of a complex must not be considered as a float-point multiplication.

The matrices of interest in the algorithm are:

❶ $M_0 = 1.\chi_0(M) - 1.\chi_6(M) - j.\chi_3(M) + j.\chi_9(M)$.

This additive matrix $M_0$ is then separated into its real and imaginary parts.

$$\Re e(M_0) = \begin{bmatrix} 1 & 1 & 1 & 1 & 1 & 1 & 1 & 1 & 1 & 1 & 1 & 1 \\ 1 & 0 & 0 & 0 & 0 & 0 & -1 & 0 & 0 & 0 & 0 & 0 \\ 1 & 0 & 0 & -1 & 0 & 0 & 1 & 0 & 0 & -1 & 0 & 0 \\ 1 & 0 & -1 & 0 & 1 & 0 & -1 & 0 & 1 & 0 & -1 & 0 \\ 1 & 0 & 0 & 1 & 0 & 0 & 1 & 0 & 0 & 1 & 0 & 0 \\ 1 & 0 & 0 & 0 & 0 & 0 & -1 & 0 & 0 & 0 & 0 & 0 \\ 1 & -1 & 1 & -1 & 1 & -1 & 1 & -1 & 1 & -1 & 1 & -1 \\ 1 & 0 & 0 & 0 & 0 & 0 & -1 & 0 & 0 & 0 & 0 & 0 \\ 1 & 0 & 0 & 1 & 0 & 0 & 1 & 0 & 0 & 1 & 0 & 0 \\ 1 & 0 & -1 & 0 & 1 & 0 & -1 & 0 & 1 & 0 & -1 & 0 \\ 1 & 0 & 0 & -1 & 0 & 0 & 1 & 0 & 0 & -1 & 0 & 0 \\ 1 & 0 & 0 & 0 & 0 & 0 & -1 & 0 & 0 & 0 & 0 & 0 \end{bmatrix}$$

which furnishes $rank(\Re e(M_0))=6$;

In SEF, the real part of the matrix is $rref \Re e(M_0)$:

$$\begin{pmatrix} 1 & 0 & 0 & 0 & 0 & 0 & 0 & 0 & 0 & 0 & 0 & 0 \\ 0 & 1 & 0 & 0 & 0 & 1 & 0 & 1 & 0 & 0 & 0 & 1 \\ 0 & 0 & 1 & 0 & 0 & 0 & 0 & 0 & 0 & 0 & 1 & 0 \\ 0 & 0 & 0 & 1 & 0 & 0 & 0 & 0 & 0 & 1 & 0 & 0 \\ 0 & 0 & 0 & 0 & 1 & 0 & 0 & 0 & 1 & 0 & 0 & 0 \\ 0 & 0 & 0 & 0 & 0 & 0 & 1 & 0 & 0 & 0 & 0 & 0 \end{pmatrix}.$$

On the other hand,

$$\Im m(M_0) = \begin{bmatrix} 0 & 0 & 0 & 0 & 0 & 0 & 0 & 0 & 0 & 0 & 0 & 0 \\ 0 & 0 & 0 & -1 & 0 & 0 & 0 & 0 & 0 & 1 & 0 & 0 \\ 0 & 0 & 0 & 0 & 0 & 0 & 0 & 0 & 0 & 0 & 0 & 0 \\ 0 & -1 & 0 & 1 & 0 & -1 & 0 & 1 & 0 & -1 & 0 & 1 \\ 0 & 0 & 0 & 0 & 0 & 0 & 0 & 0 & 0 & 0 & 0 & 0 \\ 0 & 0 & 0 & -1 & 0 & 0 & 0 & 0 & 0 & 1 & 0 & 0 \\ 0 & 0 & 0 & 0 & 0 & 0 & 0 & 0 & 0 & 0 & 0 & 0 \\ 0 & 0 & 0 & 1 & 0 & 0 & 0 & 0 & 0 & -1 & 0 & 0 \\ 0 & 0 & 0 & 0 & 0 & 0 & 0 & 0 & 0 & 0 & 0 & 0 \\ 0 & 1 & 0 & -1 & 0 & 1 & 0 & -1 & 0 & 1 & 0 & -1 \\ 0 & 0 & 0 & 0 & 0 & 0 & 0 & 0 & 0 & 0 & 0 & 0 \\ 0 & 0 & 0 & 1 & 0 & 0 & 0 & 0 & 0 & -1 & 0 & 0 \end{bmatrix}$$

which in turns yield $rank(\Im m(M_0))=2$;

In SEF, the imaginary matrix is:

$$\begin{pmatrix} 0 & 1 & 0 & 0 & 0 & 1 & 0 & -1 & 0 & 0 & 0 & -1 \\ 0 & 0 & 0 & 1 & 0 & 0 & 0 & 0 & 0 & -1 & 0 & 0 \end{pmatrix}.$$

❷ $M_1 = 1 \cdot \chi_1(M) - 1 \cdot \chi_7(M) - j \cdot \chi_4(M) + j \cdot \chi_{10}(M)$,

$$\Re e(M_1) = \begin{bmatrix} 0 & 0 & 0 & 0 & 0 & 0 & 0 & 0 & 0 & 0 & 0 & 0 \\ 0 & 1 & 0 & 0 & 0 & 0 & 0 & -1 & 0 & 0 & 0 & 0 \\ 0 & 0 & 0 & 0 & 0 & 0 & 0 & 0 & 0 & 0 & 0 & 0 \\ 0 & 0 & 0 & 0 & 0 & 0 & 0 & 0 & 0 & 0 & 0 & 0 \\ 0 & 0 & 0 & 0 & 0 & 0 & 0 & 0 & 0 & 0 & 0 & 0 \\ 0 & 0 & 0 & 0 & 0 & 1 & 0 & 0 & 0 & 0 & 0 & -1 \\ 0 & 0 & 0 & 0 & 0 & 0 & 0 & 0 & 0 & 0 & 0 & 0 \\ 0 & -1 & 0 & 0 & 0 & 0 & 0 & 1 & 0 & 0 & 0 & 0 \\ 0 & 0 & 0 & 0 & 0 & 0 & 0 & 0 & 0 & 0 & 0 & 0 \\ 0 & 0 & 0 & 0 & 0 & 0 & 0 & 0 & 0 & 0 & 0 & 0 \\ 0 & 0 & 0 & 0 & 0 & 0 & 0 & 0 & 0 & 0 & 0 & 0 \\ 0 & 0 & 0 & 0 & 0 & -1 & 0 & 0 & 0 & 0 & 0 & 1 \end{bmatrix}$$

so, $rank(\Re e(M_1))=2$; the SEF of which is

$$LI_1 := \begin{pmatrix} 0 & 1 & 0 & 0 & 0 & 0 & 0 & -1 & 0 & 0 & 0 & 0 \\ 0 & 0 & 0 & 0 & 0 & 1 & 0 & 0 & 0 & 0 & 0 & -1 \end{pmatrix}.$$

$$\Im m(M_1) = \begin{bmatrix} 0 & 0 & 0 & 0 & 0 & 0 & 0 & 0 & 0 & 0 & 0 & 0 \\ 0 & 0 & 0 & 0 & -1 & 0 & 0 & 0 & 0 & 1 & 0 & 0 \\ 0 & 0 & -1 & 0 & 0 & 1 & 0 & 0 & -1 & 0 & 0 & 1 \\ 0 & 0 & 0 & 0 & 0 & 0 & 0 & 0 & 0 & 0 & 0 & 0 \\ 0 & -1 & 0 & 0 & -1 & 0 & 0 & -1 & 0 & 0 & -1 & 0 \\ 0 & 0 & 1 & 0 & 0 & 0 & 0 & 0 & -1 & 0 & 0 & 0 \\ 0 & 0 & 0 & 0 & 0 & 0 & 0 & 0 & 0 & 0 & 0 & 0 \\ 0 & 0 & 0 & 0 & -1 & 0 & 0 & 0 & 0 & 1 & 0 & 0 \\ 0 & 0 & -1 & 0 & 0 & -1 & 0 & 0 & -1 & 0 & 0 & -1 \\ 0 & 0 & 0 & 0 & 0 & 0 & 0 & 0 & 0 & 0 & 0 & 0 \\ 0 & 1 & 0 & 0 & -1 & 0 & 0 & 1 & 0 & 0 & -1 & 0 \\ 0 & 0 & 1 & 0 & 0 & 0 & 0 & 0 & -1 & 0 & 0 & 0 \end{bmatrix}$$

and, $rank(\Im m(M_1))=6$; the SEF of which is

$$LI_2 := \begin{pmatrix} 0 & 1 & 0 & 0 & 0 & 0 & 0 & 1 & 0 & 0 & 0 & 0 \\ 0 & 0 & 1 & 0 & 0 & 0 & 0 & 0 & 0 & 0 & 0 & 0 \\ 0 & 0 & 0 & 0 & 1 & 0 & 0 & 0 & 0 & 0 & 0 & 0 \\ 0 & 0 & 0 & 0 & 0 & 1 & 0 & 0 & 0 & 0 & 0 & 1 \\ 0 & 0 & 0 & 0 & 0 & 0 & 0 & 1 & 0 & 0 & 0 & 0 \\ 0 & 0 & 0 & 0 & 0 & 0 & 0 & 0 & 0 & 1 & 0 \end{pmatrix}.$$

❸ $M_{-1} = 1 \cdot \chi_{11}(M) - 1 \cdot \chi_5(M) - j \cdot \chi_2(M) + j \cdot \chi_8(M)$

$$\Re e(M_{-1}) = \begin{bmatrix} 0 & 0 & 0 & 0 & 0 & 0 & 0 & 0 & 0 & 0 & 0 & 0 \\ 0 & 0 & 0 & 0 & 0 & -1 & 0 & 0 & 0 & 0 & 0 & 1 \\ 0 & 0 & 0 & 0 & 0 & 0 & 0 & 0 & 0 & 0 & 0 & 0 \\ 0 & 0 & 0 & 0 & 0 & 0 & 0 & 0 & 0 & 0 & 0 & 0 \\ 0 & 0 & 0 & 0 & 0 & 0 & 0 & 0 & 0 & 0 & 0 & 0 \\ 0 & -1 & 0 & 0 & 0 & 0 & 0 & 1 & 0 & 0 & 0 & 0 \\ 0 & 0 & 0 & 0 & 0 & 0 & 0 & 0 & 0 & 0 & 0 & 0 \\ 0 & 0 & 0 & 0 & 0 & 1 & 0 & 0 & 0 & 0 & 0 & -1 \\ 0 & 0 & 0 & 0 & 0 & 0 & 0 & 0 & 0 & 0 & 0 & 0 \\ 0 & 0 & 0 & 0 & 0 & 0 & 0 & 0 & 0 & 0 & 0 & 0 \\ 0 & 0 & 0 & 0 & 0 & 0 & 0 & 0 & 0 & 0 & 0 & 0 \\ 0 & 1 & 0 & 0 & 0 & 0 & 0 & -1 & 0 & 0 & 0 & 0 \end{bmatrix}$$

so, $rank(\Re e(M_{-1}))=2$; besides, the SEF of the matrix is exactly the same as the one of $\Re e(M_1)$, i.e., $LI_3=LI_1$.

$$\Im m(M_{-1}) = \begin{bmatrix} 0 & 0 & 0 & 0 & 0 & 0 & 0 & 0 & 0 & 0 & 0 & 0 \\ 0 & 0 & -1 & 0 & 0 & 0 & 0 & 1 & 0 & 0 & 0 & 0 \\ 0 & -1 & 0 & 0 & 1 & 0 & 0 & -1 & 0 & 0 & 1 & 0 \\ 0 & 0 & 0 & 0 & 0 & 0 & 0 & 0 & 0 & 0 & 0 & 0 \\ 0 & 0 & 1 & 0 & 0 & 1 & 0 & 0 & 1 & 0 & 0 & 1 \\ 0 & 0 & 0 & 0 & 1 & 0 & 0 & 0 & 0 & -1 & 0 \\ 0 & 0 & 0 & 0 & 0 & 0 & 0 & 0 & 0 & 0 & 0 & 0 \\ 0 & 0 & -1 & 0 & 0 & 0 & 0 & 1 & 0 & 0 & 0 & 0 \\ 0 & 1 & 0 & 1 & 0 & 0 & 1 & 0 & 0 & 1 & 0 \\ 0 & 0 & 0 & 0 & 0 & 0 & 0 & 0 & 0 & 0 & 0 & 0 \\ 0 & 0 & 1 & 0 & 0 & -1 & 0 & 0 & 1 & 0 & 0 & -1 \\ 0 & 0 & 0 & 0 & 1 & 0 & 0 & 0 & 0 & -1 & 0 \end{bmatrix}$$

So that $rank(\Im m(M_{-1}))=6$; Surprisingly, the SEF of this matrix is the same as the one of $\Im m(M_1)$, i.e. $LI_4=LI_2$. If fact, $\Im m(M_{-1})$ is essentially a row (or column) permutation of $\Im m(M_1)$. In order to evaluate the multiplicative complexity of the FFT of blocklength 12, we determine the rank of the matrices:

$$\Re e(M_m + M_{-m}), \text{ and } \Im m(M_m + M_{-m}),$$
$$\Re e(M_m - M_{-m}), \text{ and } \Im m(M_m - M_{-m}).$$

The four preaddition matrices associated with the multiplicative branches of the algorithm are:

$rref\, \Re e(M_1 + M_{-1}) = (0\ 1\ 0\ 0\ 0\ -1\ 0\ -1\ 0\ 0\ 0\ 1)$ (7a)

$$rref\, \Im m(M_1 - M_{-1}) = \begin{pmatrix} 0 & 1 & 0 & 0 & 0 & 1 & 0 & 1 & 0 & 0 & 0 & 1 \\ 0 & 0 & 1 & 0 & 0 & 0 & 0 & 0 & 0 & 0 & 1 & 0 \\ 0 & 0 & 0 & 0 & 1 & 0 & 0 & 0 & 1 & 0 & 0 & 0 \end{pmatrix}$$

$rref\, \Re e(M_1 - M_{-1}) = (0\ 1\ 0\ 0\ 0\ 1\ 0\ -1\ 0\ 0\ 0\ -1)$ (7b)

$$rref\, \Im m(M_1 + M_{-1}) = \begin{pmatrix} 0 & 1 & 0 & 0 & 0 & -1 & 0 & 1 & 0 & 0 & 0 & -1 \\ 0 & 0 & 1 & 0 & 0 & 0 & 0 & 0 & 0 & 0 & -1 & 0 \\ 0 & 0 & 0 & 0 & 1 & 0 & 0 & 0 & -1 & 0 & 0 & 0 \end{pmatrix}$$

For the sake of simplicity, such matrices can be put in the condensed form:

$[(0\ 1\ 0_3\ \mp 1\ 0\ -1\ 0_3\ \pm 1)]$, and

$$\begin{bmatrix} (0\,1\,0_3\ \mp 1\,0\,1\,0_3\ \mp 1 \\ 0_2\,1\,0_7\ \mp 1\,0 \\ 0_4\,1\,0_3\ \mp 1\,0_3) \end{bmatrix}.$$

Figures 1 and 2 present the block diagrams of the FFT algorithm, separating real and imaginary parts computations. The total multiplicative complexity is eight *real* floating-point multiplications, which meets Heideman's lower bound [17] and very far from the 144 multiplications required for computing the DFT by its definition.

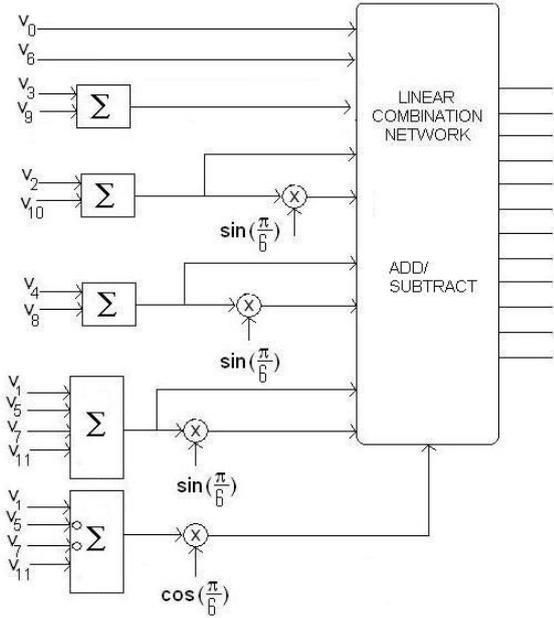

**Figure 1**. Scheme of the real part computation of a DFT ($N=12$). The small circles into the $\Sigma$-box denote subtraction. There are four multiplications for computing $\Re$ eDFT. The outputs are the twelve coefficients, which are computed by a suitable binary linear combination of inputs (Eqn 7a).

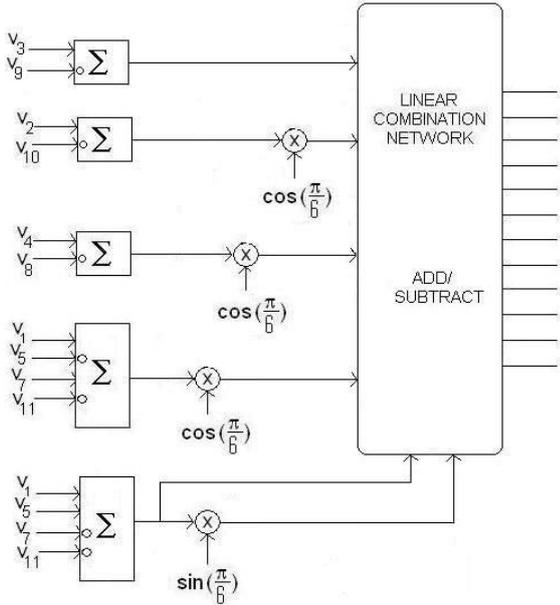

**Figure 2**. Scheme of the imaginary part computation of a DFT ($N=12$). The little circles into the $\Sigma$-box denote subtraction. The computation of $\Im$mDFT requires four floating-point multiplications. The outputs are the twelve coefficients, which are computed by a suitable linear combination of inputs (Eqn 7b). Fig.1 has similar blocks, so a joint implementation is favoured to reduce the spatial complexity of the hardware.

The complexity, in terms of real floating-point multiplications is given by $2(1+3)=8$. According with this approach, the coupled samples are:

$$v_1 \pm v_5 ; v_7 \pm v_{11} ; v_2 \pm v_{10} ; v_4 \pm v_8.$$

The presentation here was split in two figures so as to clarify the intrinsic nature of the proposed FFT algorithm: the barely required modification in the $\Re$ e-part circuit to compute the corresponding $\Im$m-part is to reverse the signal of particular input samples.

## V. COMMENTS ON THE FFT FOR FURTHER BLOCKLENGTHS

For $N=20$, there are exactly $N/4=5$ classes, corresponding to $m=0,\pm 1,\pm 2$:

$C_0=\{0,5,10,15\}$ $C_1=\{1,6,11,16\}$ and $C_{-1}=(19, 4, 9, 14)$

$C_2=\{2,7,12,17\}$ and $C_{-2}=(18, 3, 8, 13)$.

The corresponding matrices

$rref\,\Re e(M_1 + M_{-1}),\ rref\,\Re e(M_1 - M_{-1})$
$rref\,\Im m(M_1 + M_{-1}),\ rref\,\Im m(M_1 - M_{-1})$
$rref\,\Re e(M_2 + M_{-2}),\ rref\,\Re e(M_2 - M_{-2})$
$rref\,\Im m(M_2 + M_{-2}),\ rref\,\Im m(M_2 - M_{-2})$

can easily be find:

$$\begin{bmatrix} 0\,10_7 \mp 10 \text{-} 10_7 \pm 1 \\ 0_3\,10_3 \mp 10_5 \text{-} 10_3 \pm 10_2 \end{bmatrix} ; \begin{bmatrix} 0\,10_7 \mp 10 \text{-} 10_7 \pm 1 \\ 0_2\,10_{15} \pm 10 \\ 0_3\,10_3 \pm 10_5\,10_3 \pm 10_2 \\ 0_4\,10_{11} \pm 10_3 \\ 0_6\,10_7 \pm 10_5 \\ 0_8\,10_3 \pm 1\,0_7 \end{bmatrix}.$$

Table 1 presents the number of *real* floating-point multiplication required to compute the FFT for blocklengths $N \leq 60$.

Table 1. Complexity of the Laurent series-based FFT algorithm in terms of the number of *real* floating-point multiplications. Values of $N\log_2 N$ are given as a benchmark.

| $N$ | $N.\log_2 N$ (rounded) | #($N$) Laurent-FFT |
|---|---|---|
| 12 | 43 | 8 |
| 20 | 86 | 32 |
| 28 | 135 | 72 |
| 36 | 186 | 88 |
| 44 | 240 | 200 |
| 52 | 296 | 288 |
| 60 | 354 | 208 |

A comparison with Heideman's bound (Theorem 1) was not performed in Table 1 because $\mu_{DFT}$ gives the minimal number of *complex* multiplications.

Table 2. Complexity of the Laurent-based FFT algorithm in terms of the number of *real* non-trivial floating-point multiplications compared to radix-2 FFT. Rader-Brenner algorithm complexity [33] was also included.

| $N$ | $N.\log_2 N$ | Radix-2 (real nontrivial) | Rader-Brenner | Heideman-Burrus $\mu_r(N)$ | #($N$) Laurent-based FFT |
|---|---|---|---|---|---|
| 8 | 24 | 4 | 4 | 4 | 2 |
| 16 | 64 | 24 | 20 | 20 | 12 |
| 32 | 160 | 88 | 68 | 64 | 54 |
| 64 | 384 | 264 | 196 | 168 | 224 |

The fast algorithm introduced here can be used as well for any blocklength $N\equiv 0 \pmod 4$, which assures the presence of the four *eigenvalues* of the DFT, but there is no ideal symmetry in the formal series. Therefore, even though this FFT was not conceived primarily for blocklength that are a power of two [1], [31], the algorithm can also be used and complexity results are shown in Table 2, in comparison with the standard radix-2 Cooley-Tukey FFT algorithm. The Heideman-Burrus bound [32] on the minimal number of *real* multiplications needed to compute a length-$N=2^n$ DFT is $\mu_r(N) = 4N - 2\{(\log_2 N)^2 + (\log_2 N) + 2\}$. Thus, even if such lengths are not the main concern of this algorithm, the number of multiplications required by this new FFT is even below $\mu_r(N)$. However, no conflicting facts exists in here, because particular symmetries (such as $e^{-j\pi/4}$) were probably not taken into account in [32]. This is corroborated in the companion paper [34], which describes the implementation of the 16-DFT performing only

$12 < \mu_r(16) = 20$ real multiplications [34] (simulink available at URL http://www2.ee.ufpe.br/codec/Procedure_FFT.htm).

## VI. CONCLUSIONS

A new fast transform algorithm for the DFT of length $N \equiv 4$ (mod 8) is presented, which is based on symmetries of the matrices associated with a Laurent series-type development, thus providing an FFT for lengths other than the customary power of two. A naïve and illustrative instance is presented in detail for $N=12$, but the entire procedure is systematic. The multiplicative complexity of the FFT is evaluated, which achieve values less than $N \cdot \log_2 N$, for $N=12, 20, 28, 36, 44, 52, 60$. For $N=16, 32, 64$, the new FFT outperforms the best known algorithms. The arithmetic complexity (flops) of this FFT is currently under investigation. Albeit there exists scores of different and smart techniques for spectrum analysis, including the arithmetic approach [28, 29] or wavelet transforms [30], which are among the best choices, the FFTs still is an extremely widespread technique. The FFT presented here is also easy to implement using DSP or low-cost high-speed Integrated Circuits.

ACKNOWLEDGEMENTS

This work was partially supported by the Brazilian National Council for Scientific and Technological Development (CNPq) under grant#306180. The authors are deeply indebt with Gilson Silva Junior, who provided insight into the first ideas of the algorithm.